\documentstyle[epsf]{aipproc}

\newcommand{\agt}{\mathrel{\raisebox{-.6ex}{$\stackrel{\textstyle>}{\sim}$}}}
\begin{document}
\hspace*{3.7in}{\bf IUHET-417}\\
\hspace*{3.6in}{\bf December 1999}
\title{Muon Collider Physics\\ at Very High Energies\thanks{To appear in the 
{\it Proceedings of Studies on Colliders and Collider Physics at the 
Highest Energies: Muon Colliders at 10~TeV to 100~TeV}, Montauk Yacht Club 
Restor, Montauk, New York, 27 September - 1 October, 1999.}}

\author{M. S. Berger}
\address{Physics Department, Indiana University, 
Bloomington, IN 47405}

\maketitle

\begin{abstract}
Muon colliders might greatly extend the energy frontier of collider physics.
One can contemplate circular colliders with center-of-mass energies in 
excess of 10~TeV. Some physics issues that might be relevant at such a 
machine are discussed.
\end{abstract}

\section*{Introduction}
%
%
%
%

The large mass of the muon compared to that of the electron results in 
a large suppression of bremstrahlung radiation. Consequently it is possible
to consider building circular colliders with energies in the multi-TeV 
regime\cite{muon}. Muon colliders have been proposed as Higgs factories and 
more recently as neutrino factories, but the long-term goal of muon colliders
should be to extend the energy frontier. It is not clear
at the present time whether advances in accelerator technology will 
result in electron-positron machines achieving
energies of several TeV. In this workshop first attempts were made to explore
the feasibility of muon colliders with energies of at least 10~TeV.

It is hard to know what kind of physics might present itself in the 10-100~TeV
mass range. After all, physicists have been arguing for a long time about the 
physics that will manifest itself at the Large Hadron Collider (LHC). The LHC,
linear electron-positron colliders, and
perhaps muon colliders should give us some clue as to what to expect at 
the following generation of machines. It is easy to imagine 
scenarios where a new collider might be necessary, but it is impossible to 
motivate a specific energy at this time. We can only speculate as to what 
physics might appear at the LHC or future linear colliders.

\section*{Luminosity requirements}

The figure of merit for physics searches at a muon collider is the QED cross
section $\mu^+\mu^-\to e^+e^-$, which has the value
\begin{eqnarray}
&&\sigma _{QED}^{}={{100~{\rm fb}}\over {s~({\rm TeV}^2)}}
\end{eqnarray}
To arrive at a simple estimate of the integrated luminosity needed to study
new physics, we assume 
\begin{eqnarray}
&&\left( \int {\cal L} dt \right) \sigma_{QED} \agt 1000\rm\ events 
\end{eqnarray}
Then the luminosity requirement for this number of events to be accumulated
in one year's running is
\begin{eqnarray}
&&{\cal L} \agt 10^{33} \cdot s\rm\ (cm)^{-2} \, (sec)^{-1} \nonumber
\end{eqnarray}
For the colliders with the center-of-mass energies considered at this meeting:

\begin{itemize}

\item $\sqrt s \simeq 10$ TeV, requiring
\begin{eqnarray}
&&\int {\cal L}dt \agt 1\rm\ (fb)^{-1}, \quad
{\cal L} \agt 10^{35}\rm\ (cm)^{-2}\, (sec)^{-1}\nonumber
\end{eqnarray}

\item $\sqrt s \simeq 100$ TeV, requiring
\begin{eqnarray}
&&\int {\cal L}dt \agt 100\rm \ (fb)^{-1}, \quad
{\cal L} \agt 10^{37}\rm\ (cm)^{-2}\, (sec)^{-1}\nonumber
\end{eqnarray}

\end{itemize}
These luminosities are extremely high, of course, and it is not clear if 
experiments can be performed in such an environment.

\section*{Electroweak symmetry breaking}

A 10~TeV muon collider might be very useful for exploring the physics 
responsible for electroweak symmetry breaking.
If Higgs bosons with $m_H<{\cal O}(800)$ GeV do not exist then interactions 
of longitudinally polarized weak bosons $(W_L,Z_L)$ become strong and can
be probed by studying vector boson scattering as shown in the figure. 
Therefore,
new physics must be present at the TeV energy scale. While one can study 
strong $W_LW_L$ scattering at the LHC, linear colliders, or $\mu^+\mu^-$ colliders with a few TeV center-of-mass energy, it might become necessary to go
to higher energies to fully explore the multitude of resonances. Indeed
we are still studying the analogous spectrum of QCD today.

\vspace*{-5.5cm}
\epsfxsize=12cm
\hspace{1cm}\epsfbox{feynman.ps}
\vspace*{-5.5cm}

\section*{Fermion mass generation}

The mechanism responsible for fermion masses and the mechanism breaking the 
electroweak symmetry are the same in the Standard Model. A Higgs scalar
acquires a vacuum expectation value giving rise to massive gauge bosons and
(through Yukawa couplings) masses for the fermions. However, it need not be
the case that these mechanisms are the same, and technicolor models were the
most prominent examples of theories where the fermion masses arise from a 
different sector from that responsible for the electroweak symmetry breaking. 
Hence one should keep an open mind about the origin of fermion masses. 
Very general 
constraints one can place on the physics of fermion mass generation are
unitarity bounds. The relevant bound for fermions scattering into 
longitudinally polarized vector boson $V_L$,
\begin{equation}
f\overline{f}\to V_LV_L\;,
\end{equation}
is the 
Appelquist-Chanowitz bound\cite{ac} which states that unitarity is violated
at the scale 
\begin{equation}
\Lambda_f<{{8\pi v^2}\over {\sqrt{3N_c}m_f}}\;,
\end{equation}
where $v=(\sqrt{2}G_F)^{-1/2}$ is the electroweak vev and $N_c$ is the number
of colors of the fermion. In the Standard Model this unitarity violation is
cured by the inclusion of the $s$-channel Higgs exchange diagram. The strongest
bound comes for the heaviest fermion the top quark for which 
$\Lambda _t\approx 3$~TeV, indicating that some new physics must occur
below this scale.   
 
For a muon one gets $\Lambda _\mu\approx 8,000$~TeV. So if the physics 
responsible for the muon mass saturates this bound, it is beyond the reach
even of a 10-100~TeV muon collider. But one does not really expect that the 
bound is saturated, but rather that the fermion masses are all generated at
a common scale with some masses suppressed by some approximate flavor 
symmetries. In light of the lower value of $\Lambda _t$, one might expect a 
10~TeV collider to provide important insight into fermion mass generation if 
Nature is not so kind to provide a elementary scalar particle.
In the typical case one expects the resonances to be broad. In some
scenarios\cite{be}, one can have strongly 
interacting Higgs sectors with narrow resonances for which a 
small energy spread might be helpful.

One can also study the unitarity violation in the subprocess 
$V_LV_L\to t\overline{t}$, analogous to the case discussed in the 
previous section for electroweak symmetry breaking. This process could also
be sensitive to new physics responsible for the fermion masses, and one would 
measure the cross sections
for $\mu^+\mu^-\to \nu\overline{\nu}t\overline{t}$ and 
$\mu^+\mu^-\to \mu^+\mu^-t\overline{t}$ , and in scenarios where the 
unitarity is saturated, one might need the energy reach of a very high 
energy muon collider to probe these strong interactions.

\section*{Gauge Bosons}
A favorite target for new physics is the possibility of new gauge bosons 
beyond those found in the Standard Model. One might first reveal the
existence of these particles via radiative return\cite{rad} whereby a vector
boson with mass less than the center-of-mass energy is produced in 
association with an energetic photon. Alternatively one could pinpoint the 
mass of the vector boson by doing precision measurements of the couplings
and asymmetries at energies below the vector boson mass. In either case, one
would ultimately want to build a collider with an energy equal to the mass of
the vector boson and take advantage of the resonance cross section. 
An important consideration then is the beam energy spread of the muon collider.
The width of the vector boson should scale linearly with its mass. 
The expectations for a 10~TeV collider is that the energy spread $\sigma_E/E$
should be something like $10^{-4}-10^{-3}$\cite{bking}, 
so the spread should be 
much smaller than the resonance peak in the typical case.

\section*{Supersymmetry}
It is possible that the LHC and linear colliders will uncover only part of the 
supersymmetric (SUSY) spectrum. 
In fact the lightest two generations of squarks and
sleptons might appear at the multi-TeV scale. The absence of certain 
supersymmetric partners being produced below the TeV energy scale would 
certainly compel us to go to higher energies.

Beyond the discovery of all the superpartners to the Standard Model particles,
another possible role for a very high energy muon collider would be to uncover
an entirely new sector responsible for the dynamical breaking of
supersymmetry. In gravitationally mediated SUSY breaking, the dynamical sector
is hidden and couples only via gravitational couplings to the supersymmetric
Standard Model particles. However other scenarios of SUSY breaking are 
possible, and these can be directly probed with sufficiently energetic 
collisions. In gauge mediated SUSY breaking scenarios, 
for example, there is just such another sector (known as the messenger sector)
occurs
at a scale beyond that which can be probed at the LHC. This messenger
sector might perhaps be 
accessible at a very high energy muon collider. The LHC might indirectly 
provide clues about the source of SUSY breaking by measuring the spectrum of 
superpartners and perhaps seeing radiative decays in the case of gauge
mediated SUSY breaking. In fact by measuring the location of displaced 
vertices (relative to the interaction point) from
the radiative decay of the next-lightest supersymmetric particle one can 
put a constraint on the scale of the gauge mediation sector as 
first suggested in a Very Large Hadron Collider study\cite{vlhc}.   

\section*{Compton Backscattering}
It seems at first peculiar to consider backscattering photons off of a muon 
beam. After all, the reason to employ muon beams rather than the electron beams
is to decrease electromagnetic radiation. Eventually however, even for muons, 
bremstrahlung radiation would again become a 
problem at sufficiently high energies in a circular collider.
At the energies contemplated here, one can reconsidering employing 
Compton backscattering to produce photon beams of comparable energies.
Kinematics dictates that the highest energy of a backscattered photon that 
can be obtained is given by 

\begin{equation}
\omega_{\rm max}={{x}\over {1+x}}E_{\rm beam}\;,
\end{equation}
where
\begin{equation}
x={{4E_{\rm beam}\omega_{\rm laser}}\over {m^2_\mu}}\;.
\end{equation}

Assuming an incident laser with energy $1.17$~eV\footnote{For definiteness we 
take a neodinium glass laser with $\omega_{\rm laser}=1.17$~eV which is 
often considered for Compton scattering at a linear $e^+e^-$ collider. 
In any case, one expects the laser energy to be in the few eV range.}, 
one obtains maximum 
backscattered photon energies (shown in the figure) which are still much 
smaller than the incident muon beam energy. A more energetic photon source
would be needed to fully realize the backscattered photon option even at 
the extremely high muon energies considered here.

\vspace*{0.0cm}
\epsfxsize=10cm
\hspace{1cm}\epsfbox{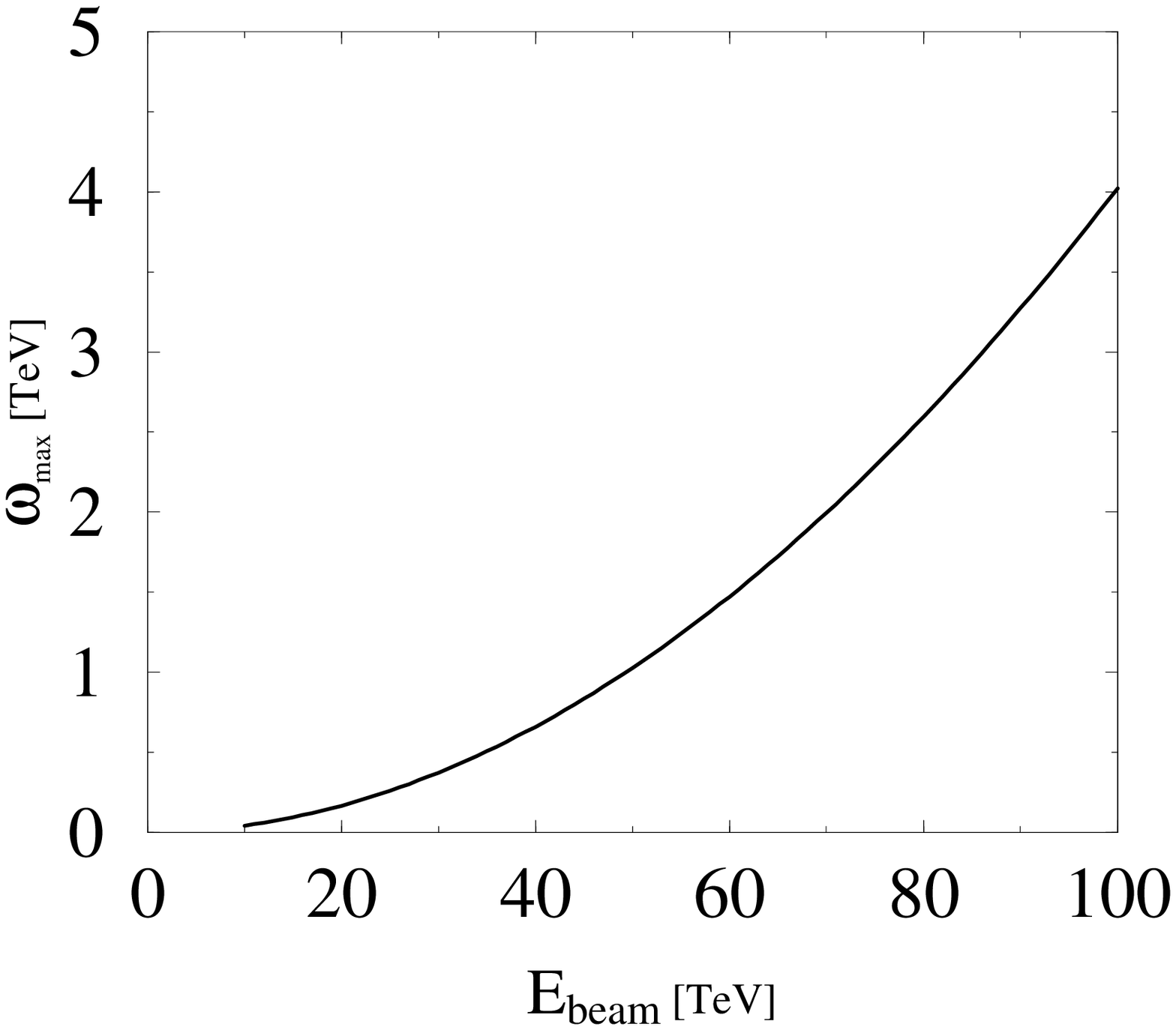}
\vspace*{0.0cm}

\section*{Conclusions}
It is difficult to motivate a very high energy muon collider without 
information that will be gleaned after years of operation of the LHC and 
linear colliders. However, if the past history of particle physics has taught
us anything it is that the most important progress has occurred by going to 
higher and higher energies. It will be interesting in the coming years 
to learn whether multi-TeV muon colliders are realistic and economical.
   
\section*{Acknowledgement}
Work supported in part by the U.S. Department of
Energy under Grant No. DE-FG02-95ER40661.

\end{document}